\title{\bf The Phase Diagram of the $U(2)\times U(2)$ Sigma Model\\
           and its
            Implications for Chiral Hierarchies.}

\author{ {{\sc D.Espriu}\thanks{espriu@greta.ecm.ub.es},\
           {\sc V.Koulovassilopoulos}\thanks{vk@tholos.ecm.ub.es} \
           and \
           {\sc A.Travesset}\thanks{alex@greta.ecm.ub.es}}\\
	 {Departament d'Estructura i Constituents
		de la Mat\`eria} \\
	 { Universitat de Barcelona} \\
	 { and} \\
	 { Institut de F\'\i sica d'Altes Energies} \\
	 { Diagonal, 647} \\
	 { E-08028 Barcelona}\\
	 \\ }

\date{}
\documentstyle[11pt,epsf]{article}

\newcommand{\be}{\begin{equation}}
\newcommand{\ee}{\end{equation}}
\newcommand{\bea}{\begin{eqnarray}}
\newcommand{\eea}{\end{eqnarray}}
\newcommand{\der}[2]{\frac{\partial #1}{\partial #2}}

\newcommand{\bref}[1]{(\ref{#1})}
\newcommand{\vr}{\varphi}
\newcommand{\ha}[1]{\cal{#1}}
\newcommand{\aam}{{\overline m}^2}

\begin{document}
 \maketitle
\begin{abstract}

Motivated by the issue of whether it is possible to construct
phenomenologically viable models where the electroweak symmetry breaking
is triggered by new physics at a scale $\Lambda \gg 4\pi v$, where $v$ is
the order parameter of the transition  ($v\sim 250$ GeV) and
$\Lambda$ is the scale of new physics, we have
studied the phase diagram of the $U(2) \times U(2)$  model. This
is the relevant low energy effective theory for a class of models which
will be discussed below. We find that the phase transition in these
models is first order in most of parameter space. The
order parameter can not be made much smaller than the cut-off and,
consequently a large hierarchy does not appear sustainable. In
the relatively small region in the space of parameters where the
phase transition is very weakly first order or second order the
model effectively reduces to the $O(8)$ theory for which the triviality
considerations should apply.

\end{abstract}

\vfill
\vbox{
UB-ECM-PF 97/04\null\par
May 1997}

\clearpage

\section{Introduction}\label{int}

In the  minimal version of the Standard Model of electroweak interactions
the same mechanism (a one-doublet complex scalar field) gives
masses simultaneously to the $W$ and $Z$ gauge bosons and to the
fermionic matter fields (other than the neutrino). This remains so in
many extensions of the minimal Standard Model, such as those consisting
of more scalar doublet fields or even fields in other representations
of $SU(2)_L$.

On the contrary, the mechanism that gives masses to the $W$ and $Z$ bosons
and to the matter fields remains somewhat distinct in models of
dynamical symmetry breaking (such as technicolor theories \cite{TC}).
In these models, there are interactions that become strong,  typically
at the scale $4\pi v$ ($v=250$ GeV), breaking the global $SU(2)_L\times
SU(2)_R$ symmetry to its diagonal subgroup $SU(2)_V$ and producing
Goldstone bosons which eventually become the longitudinal degrees of
freedom of the $W^\pm$ and $Z$.
Yet, to transmit this symmetry breaking to the ordinary matter fields
one usually requires  additional interactions, characterized by a
different scale $M$. Generally, it is assumed that $M\gg 4\pi v$.
It seems then natural to ask whether it is necessary to have these two
very different scales and  whether it would not have been possible to
arrange things in such a way that the $SU(2)_L\times SU(2)_R\to SU(2)$
symmetry  breaking  takes place at a scale $\Lambda$, where $4\pi v \ll
\Lambda $. Although the scale where the symmetry breaking
takes place, $\Lambda$, and the scale characterizing
the new physics, $M$,
need not be exactly the same, we shall assume that $\Lambda\simeq
M$ and refer only to $\Lambda$ hereafter.

Two phenomenologically viable paradigms of the above possibility are the
strong-coupling extended technicolor (ETC) models \cite{SETC}, and the
top-condensate (TopC) models \cite{TOPC}, in which the underlying
dynamics is, typically, a spontaneously broken gauge theory,
characterized by a scale $\Lambda$, with $\Lambda \gg 1$~TeV.
At sufficiently low energies, the dynamics can be modeled by four-fermi
interactions (of either, technifermion doublets in ETC models, or the
quarks of the third family in TopC models) which are attractive in the
scalar channel. Then, it  appears possible that, by tuning the corresponding
coupling sufficiently close, but above a critical value,  chiral
symmetry breaking occurs, but the condensate itself is of the
order of the weak scale $v$, much smaller than its natural value
${\cal O}(\Lambda)$.
It has been pointed out in  ref.~\cite{CGS}, that a necessary condition
for this to happen is that the low-energy effective theory, which
retains the light degrees of freedom below the scale $\Lambda$,
where chiral symmetry breaking occurs,
possesses itself a second order phase transition. It is only then that
there can consistently be a hierarchy between the order parameter $v$
and the scale $\Lambda$.

If the strongly interacting fermions at scale $\Lambda$ are electroweak
doublets, then the chiral symmetry is $U(2)_L\times U(2)_R$ and the
relevant lagrangian at that scale consists in a bunch of four-fermion
interactions. The precise form of these four-fermi interactions will not
concern us here. It has been argued, using analytical methods \cite{CCL}
as well as lattice simulations \cite{HASENFRATZ,SMIT}, that four-fermion
models are, at low energies, equivalent to an effective theory
consisting in a scalar-fermion model with the appropriate symmetry,
which in our case it will be $U(2)_L\times U(2)_R$.

Using an effective theory description also frees us from
having to appeal to any particular model. Thus, the analysis
may remain valid beyond the models we have just used to motivate the
problem.
In this effective theory we need to  retain only the particles
that remain light after chiral symmetry breaking.
Thus, it must necessarily contain the Goldstone bosons
emerging from the breaking of the global $U(2)_L\times U(2)_R$ symmetry.
There may also be some light (compared to the scale $\Lambda$)
scalars. If present, the $U(2)_L\times U(2)_R$ symmetry
can be linearly realized. If not, the symmetry should be realized
non-linearly. Of course the presence or not of such
additional scalars depends only on the underlying additional
sector (or, equivalently, on the four-fermion interactions
it leads to). However, the linear and non-linear theories
differ, for sufficiently large values of the scalar masses,
 by terms of ${\cal O}(
\mu^2/16\pi^2 v(\mu)^2)$. In the coming paragraphs
we just argue that $v$ is generally large. Hence,
at low energies these terms are small and thus using a linear
realization is really no restriction at all.

We shall thus
assume that the low-energy theory is a general
linear sigma model with $U(N)_L\times U(N)_R$ (which we later take
$N=2$) symmetry, whose effective action, for general $N$, is given by
\bea
S(\phi)&=&\int d^4x (\frac{1}{2}{\rm Tr} (\partial_{\mu}\phi^{\dagger}
\partial_{\mu}\phi)+
\frac{1}{2}m^2{\rm Tr} ( \phi^{\dagger} \phi)
% \nonumber
% \\ &&
+\lambda_1 ({\rm Tr} \phi^{\dagger} \phi)^2+
\lambda_2 {\rm Tr} ( \phi^{\dagger} \phi)^2 ),
\label{action}
\eea
where $\phi(x)$ is a complex $N \times N$ matrix, the order parameter of
the high-energy phase transition. The action \bref{action} is invariant
under the global symmetry transformation $\phi \rightarrow L \phi
R^\dagger$,  where $L,R$ are $U(N)$ matrices. The electroweak
interactions are obtained by gauging an $SU(2) \times U(1)$ subgroup of
this global symmetry,  which after the $U(N)_L\times U(N)_R$ symmetry
breaking gives masses to the $W^\pm$ and $Z$.
In some models additional fermions remain in the spectrum below
$\Lambda$. We have not considered them and thus our results
do not apply to such models.

This model, as will be discussed in detail below, possesses for
$\lambda_2 \neq 0$ a first order phase transition whose strength varies
in the $(\lambda_1, \lambda_2)$ space. As one adjusts $m^2$ past a
critical value ($m^2=0$ in mean field theory), the vacuum expectation
value $v$ jumps discontinuously from zero in the unbroken phase to
some finite nonzero value in the broken phase.
If the couplings $(\lambda_1, \lambda_2)$, which are
obtained by matching with the underlying strong dynamics at the cut-off
$\Lambda$, belong to a region where the phase transition is
weakly first order,
then the vacuum expectation value (v.e.v.) $v$ can be small, $v/\Lambda
\ll 1$, and the $U(2)_L\times U(2)_R$ model can still be a valid
description of the low-energy dynamics. 
If, on the other hand, the couplings $(\lambda_1, \lambda_2)$ belong to
a region where the transition is strongly first order, such a hierarchy
is not possible and models leading to such $(\lambda_1,\lambda_2)$
values should be excluded.

It is our purpose in this paper to investigate in detail the phase
structure of the above model, and to conclude whether large chiral
hierarchies are sustainable in this context.  A number of authors have
previously considered this possibility.
In \cite{CGS} such a model was considered as a possible parametrization
of the low-energy physics of top-condensate models or strongly coupled 
extended technicolor models. The authors concluded that, within the
framework of a perturbative one-loop analysis, a large hierarchy is
unlikely unless $\lambda_2$ is close to zero. 
Later, in \cite{BHJ}, it was pointed out that a leading order $1/N_c$ 
analysis combined with two-loop beta functions can change the
conclusions, and that consistent large hierarchies were not
disallowed. Unfortunately, all these analysis rely on perturbation
theory which is unreliable at strong coupling. To settle the issue we have
performed Monte Carlo simulations in terms of the lattice regularized
version of the action (\ref{action}) above.
A preliminary investigation along these lines was undertaken in
ref.~\cite{YUE}.

We confirm the first-order character of the transition for $\lambda_2
\neq 0$. We have obtained a detailed picture of the behaviour of the
order parameter for nonperturbative values of the sigma model
couplings and semi-quantitative estimates of the  correlation
length. Where meaningful we compare our results to those
obtained via the effective potential. We have found that a large
hierarchy is untenable in most of parameter space. $v$ is typically
several orders of magnitude too large.

To get  the above results we  have performed high-statistics runs using
a hybrid algorithm  (with and without Fourier acceleration) which, to
our knowledge, had not been used before for this type of systems. We have
also written a traditional Metropolis  Monte Carlo code for comparison.
Therefore we believe that our results are also of some interest to the
lattice expert.

\section{The $U(2)_L\times U(2)_R $ model}

In the case $N=2$
the action (\ref{action}) depends on eight degrees of freedom and the
field $\phi$ can be
conveniently parametrized by
\be
\phi=\sum_{a=0}^{3} (\sigma^a+i \pi^a)\frac{\tau^a}{\sqrt{2}}
\ee
where $\tau^a$ are the Pauli matrices for $a=1,2,3$, and the identity
matrix for $a=0$.
The action (\ref{action})  is  invariant under the rigid symmetry, $\phi
\rightarrow L \phi R^{\dagger}$, where $L,R \in U(2)$. If we set
$\lambda_2=0$, then the symmetry is enhanced to $O(8)$ ($O(2N^2)$ for
general $N$).

The pattern of symmetry breaking depends on the sign of the coupling
$\lambda_2$. If $\lambda_2>0$ the vacuum expectation value (v.e.v.) can be
rotated to $<\sigma_0>=v$, and the breaking occurs according to
\be
U(2)_L \times U(2)_R \rightarrow U(2)_V
\ee
with  $v^2=-m^2/(4\lambda_1+2\lambda_2)$. The $\pi^\alpha$ are then
the Goldstone bosons while the masses for the other states are
\be
m_{\sigma_0}^2=4(2\lambda_1+\lambda_2) v^2 \;\; , \;\;\;\;
m_{\sigma_i}=4\lambda_2 v^2 , \;\;\; i=1,2,3.
\ee
If $\lambda_2 <0$, the symmetry breaks along the $\tau^0+\tau^3$
direction,
\be
U(2)_L \times U(2)_R \rightarrow U(1)^3 \label{SB2}
\ee
with $v^2=-m^2/2 (\lambda_1+\lambda_2)$, and the masses are
\be
m_{\sigma_0}^2=m_{\pi_0}^2=-2\lambda_2 v^2 \;\; , \;\;\;\;
m_{\sigma_3}=4(\lambda_1+\lambda_2) v^2 .
\ee
In this case though, the mean field solution is not a real minimum but a
saddle point.

Although these are tree-level relations, they are preserved by quantum
corrections in terms of renormalized parameters appropriately defined.
In the physically interesting region, and barring any unexpected 
non-trivial fixed point, four dimensional scalar theories such as in
(\ref{action}) are believed to be infrared free and have Landau poles in
the ultraviolet.
Therefore if we wish to have large masses for all these scalar
resonances we must set at least $\lambda_2(\Lambda)\gg 1$. (Recall that
the bare coupling and the renormalized coupling at the cut-off scale are
the same thing.). Let us keep in mind, though, that even after
giving these scalars a large mass, some non-decoupling effects
remain, as pertain to a spontaneously broken theory.

The electroweak interactions are included by identifying
$SU(2)_L$ with $SU(2)_W$ and the $\tau_3$ component of $SU(2)_R$ with
hypercharge. The symmetry of the model is expected to break according
to $U(2)_L\times U(2)_R \to U(2)_V$, producing four Goldstone
bosons, with the vacuum expectation value related to the weak scale
via $M_W=gv/2$. Then three Goldstone bosons become the longitudinal
components of the $W^{\pm}$ and $Z$ bosons, while the fourth one is
expected to get a mass from the anomalous breaking of the axial $U(1)$.
Physical fermions that do not feel the strong interactions and, hence do
not participate in the symmetry breaking, should remain light due to
their small Yukawa couplings,  $y$ and, will be ignored henceforth.
Similarly, we have neglected the effects of gauge couplings since they
are small at scale $\Lambda$. We have also ignored the possible
presence of custodially non-invariant interactions.

As discussed, a non-perturbative method, such as lattice techniques, 
is called for to determine the possibility of making $v$ of the order of
weak scale. 
After introducing the lattice regulator, one expects that $\Lambda/v
\sim \xi^{\beta}$, with $\beta$ the appropriate critical exponent, so a
hierarchy will only be possible if the correlation length $\xi$ is big,
or what is the same, the transition is second order or weakly first
order.

High-energy physicists have not devoted to first order phase transitions
nearly as much interest as in the case of continuous ones since there is
no natural way to define a continuum limit, i.e. to shrink the lattice
spacing to zero, because the correlation length never becomes infinite.
However, this is not a problem from the point of view of an effective
field theory
because our continuum theory has most definitely an
ultraviolet cut-off $\Lambda$, above which it is
no longer valid.
The lattice cut-off can then be identified with this continuum
ultraviolet cut-off, i.e. $a=2\pi\Lambda^{-1}$. The relation between the
lattice and the continuum cut-off can be unambiguously established, but
since we will not work it out here all we can say is that the relation
between the lattice cut-off and the physical scale $\Lambda$ can be
defined up to terms of ${\cal O}(\Lambda^{-2})$ only.
The physical parameter controlling the size of the corrections is
naturally the correlation length of the system.
If the correlation length is relatively large (in lattice units),
corrections will be small, non-universal cut-off effects controlable,
and the results meaningful.
It turns out that in most of the interesting regions of parameter space,
the transition is of first order, but with relatively large
correlation lengths.
We can thus deposit some confidence in our conclusions.

\section{The Coleman-Weinberg mechanism}

In this section, we analyze the phase transition within
(renormalized) perturbation theory. Although this applies strictly only
at weak coupling, that will provide a qualitative feeling about the
transition. Moreover, it will be suggestive of the regions in coupling
constant space where we should perform Monte Carlo simulations and
provide a qualitative understanding of our results.

The model described by (\ref{action}) is very similar to the
Ginzburg-Landau phenomenological model of continuous
transitions. However, theoretical considerations  \cite{PATERSON} and
numerical work show that, whenever $\lambda_2 \neq 0$, by tuning the
parameter $m^2$ ($m^2<0$) the system undergoes a first-order transition,
which particle physicists know as the Coleman-Weinberg mechanism
\cite{CW}.  Due to quantum fluctuations the system develops a vacuum
expectation value at a finite value of the correlation length $\xi$.

The Coleman-Weinberg mechanism has been given a nice geometrical
interpretation  in a massless theory, due to Yamagishi \cite{YAM}, in
terms of the $\beta$ functions, and its associated fixed points.
The general solution of the renormalization group equation (in a
dimensionless regulator, such as dimensional regularization) for the
effective potential $V(\varphi)$
\be
(\mu{\partial\over{\partial\mu}}+\beta_1{\partial\over{\partial\lambda_1}}
+\beta_2{\partial\over{\partial\lambda_2}}
+\gamma \varphi{\partial\over{\partial\varphi}})V(\varphi)=0
\label{rgeq}
\ee
where $\varphi= ({\rm Tr}\phi^\dagger \phi)^{1/2}$, when $\lambda_2>0$
is given by
\be
V(\varphi)=
\left(\lambda_1(t)+\frac{\lambda_2(t)}{2}\right) \varphi^4\times
\exp\left(4\int_0^t dt { \gamma(t)\over{ 1-\gamma(t)}}\right)
\label{solution1}
\ee
where $t=\ln(\varphi/\mu)$. Then, the condition for the existence of a
local extremum away from the origin $\langle\varphi\rangle=v\neq 0$
leads to
\be
4 (2\lambda_1(t)+\lambda_2(t))+2\beta_1+\beta_2=0
\label{stability1}
\ee
where $\beta_i=\partial\lambda_i/\partial t, i=1,2$
are the $\beta$-functions, with initial conditions
$\lambda_1(0)=\lambda_1, \lambda_2(0)=\lambda_2$. Eq.~(\ref{stability1})
is referred to as the ``stability line''.

The corresponding RG equation for $\lambda_2<0$ is given by
\be
V(\varphi)=
(\lambda_1(t)+\lambda_2(t)) \varphi^4\times
\exp\left(4\int_0^t dt { \gamma(t)\over{ 1-\gamma(t)}}\right)
\label{solution2}
\ee
% where  $\varphi=({\rm Tr}\phi^\dagger \phi)^{1/2}$. Then
and the stability line is described by
\be
4 (\lambda_1(t)+\lambda_2(t))+\beta_1+\beta_2=0   .
\label{stability2}
\ee
If there exists a certain value  of $t$ where the conditions
(\ref{stability1}) or (\ref{stability2}) are satisfied, in a region
where $V^{\prime\prime}>0$ and $V<0$ at the minimum (see for the
corresponding equations in terms of the appropriate
$\beta$ functions in
ref.~\cite{YUE})  then $\langle\varphi\rangle=v\neq 0$ and the
transition is of first order.

\begin{figure}[htb]
  \epsfxsize 18cm \centerline{\epsffile{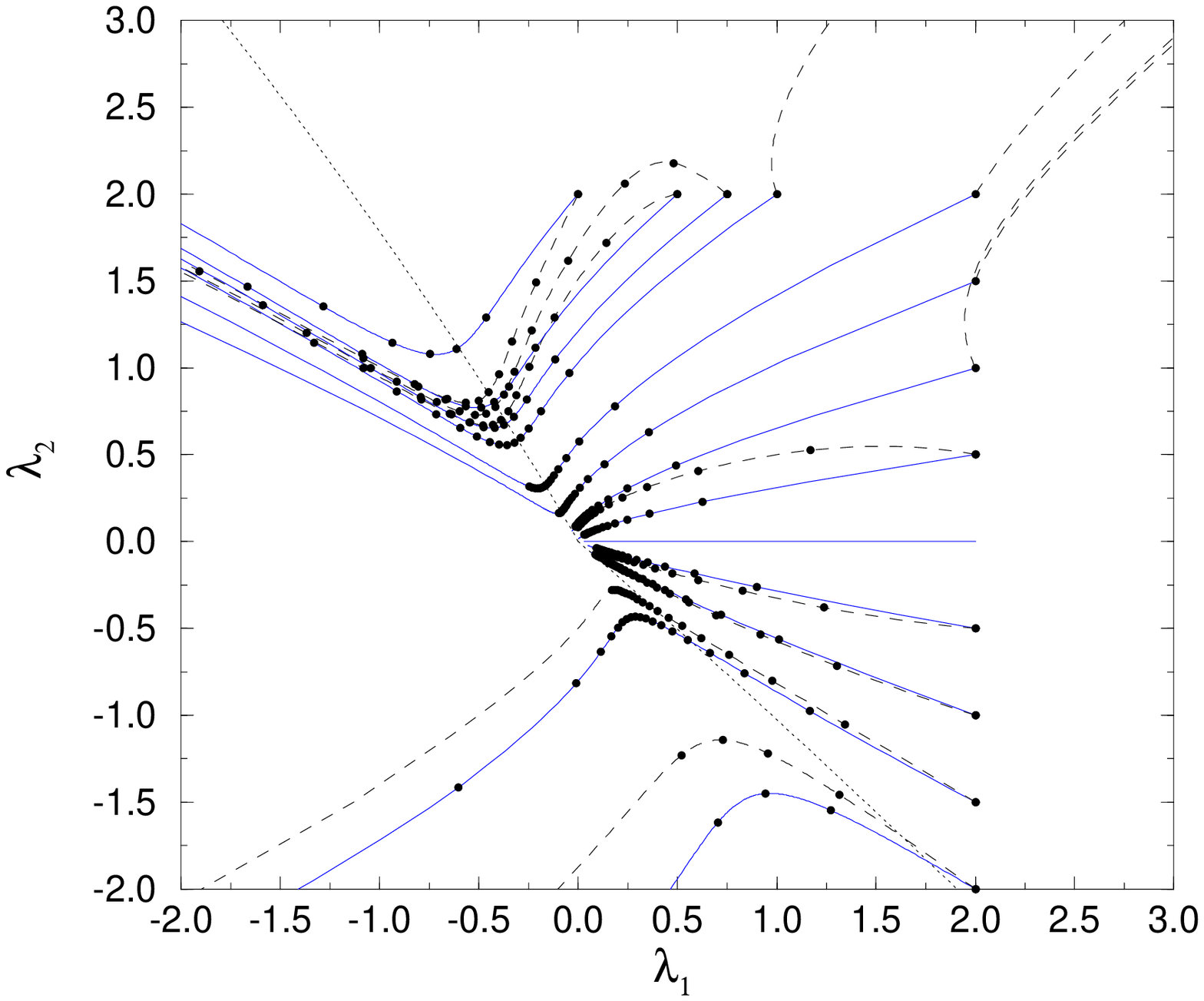}}
\caption{}
 \label{traj}
\end{figure}

The expressions for the $\beta$ functions can be found at one-loop level
in \cite{PISA,CGS,YUE} and are plotted as solid lines in fig.~(\ref{traj}).
The stability line is indicated as a dotted line. Then, starting from
some value $(\lambda_1, \lambda_2)$  at the scale 
$\Lambda$ and following its RG trajectory,  one flows in the infrared
either towards the stability line or towards the infrared fixed point at
the origin (if $\lambda_2=0$). If the RG trajectory crosses the
stability line, then the transition must necessarily be of first order
at that particular value of $(\lambda_1,\lambda_2)$ we started. Were it
of second order, the correlation length would be divergent and it cannot
possibly become finite again after a finite number of renormalization group
blockings. For $|\lambda_2|$ small the couplings flow towards
the region $\lambda_1,\lambda_2 \ll 1$ and even if they cross the
stability, they do so after very many decades of running; the phase
transition in this case is weakly first order, the more so as
$|\lambda_2|\rightarrow 0$.

The corresponding $\beta$ functions at two-loop level can be found in 
ref.~\cite{MACHACEK,BHJ}, whose solutions are plotted (for zero Yukawa
coupling) in  fig.~(\ref{traj}) with dashed lines.
One finds out that the stability is improved by the two-loop corrections
\cite{BHJ}. For a bare theory  with $\lambda_2<0$ or one that is
close to the stability line the flow is again to the left towards the
stability line, however the flow is slower than at one-loop. 
Furthermore, there exists a region with $\lambda_1,\lambda_2>0$  where
the flow is reversed and it appears that it never crosses the stability
line.  However, this only hints upon the breakdown of perturbation
theory and a nonperturbative analysis is called for. 

Although strictly valid only with a dimensionless regulator, and hence
definitively linked to perturbation theory, the conclusions of the above
analysis are expected to remain approximately valid in the lattice
regularization where all sorts of irrelevant operators appear in the
effective potential. As long as the correlation length is large
enough, the continuum physics can be used as a guidance.
Checking to what extend these arguments are valid is one of the motivations
of the present work.

It is also essential that no other fixed point unreachable in
perturbation theory exists. Should one be present, the RG trajectories
would be distorted and there could be
regions where the transition is second order. We found no evidence of
such a fixed point. Even with only the gaussian fixed point it would
still be conceivable that there might exist a region of non-zero
measure whose RG trajectories end in the gaussian fixed point at the
origin.  For these values the transition would be of second order.  For
small values of ($\lambda_1, \lambda_2$) an effective potential
calculation shows that this happens only if  $\lambda_2=0$, so assuming
that the RG trajectories follow a potential flow this possibility
appears to be ruled out too.

The above picture suggests that, if the couplings of a
given bare theory are located in the region limited by the
stability line
and the straight lines
$2\lambda_1(\Lambda) +\lambda_2(\Lambda)=0$ for $\lambda_2>0$ and
$\lambda_1(\Lambda) +\lambda_2(\Lambda)=0$ for $\lambda_2<0$
(so that the
potential is positive definite at large $\phi$), then one should observe
a first order transition when $m^2$ crosses the critical surface.
On the other hand, if the renormalized couplings
are located to the right of the stability line, first order transitions
should also be observed near the stability line, decreasing in
strength as we separate from it and also with decreasing $\lambda_2$ as
the correlation length increases when we approach the $\lambda_2=0$
line.

It should be emphasized here that the situation here
is different from the triviality analysis in the $O(N)$ model
(i.e. $\lambda_2=0$ axis) \cite{TRIVIALITY}. The phase transition there
is of second order and all points belong to the attraction of the
gaussian fixed point, where the continuum limit is just a free
theory. However, for given quartic couplings, one can 
always define a consistent effective field theory, with an
arbitrarily large hierarchy $\Lambda/v$, at the expense though of
having an upper bound on the masses of the scalar particle.
If the mass is of the order of the  cut-off,
lattice artifacts creep in and
we cannot really consider the model as as a field theory one.
The situation is, on the other hand, different for the $U(2)_L\times
U(2)_R$ model, since the gaussian fixed point is infrared unstable:
unless $\lambda_2=0$, the renormalization group flows do not belong to
its attractive domain, and, in the absence of another fixed point,
become runaway trajectories. In this case, there is no proper continuum
limit and strictly speaking no field theory at all. 
However, if the transition is sufficiently weakly first order one
can speak of an approximate continuum limit, with lattice artifacts
being still relatively small. In contrast to the  $\lambda_2=0$ case
though,  the correlation length is not tunable (by $m^2$) but is rather
determined by the quartic couplings ($\lambda_1, \lambda_2$). Moreover,
a large hierarchy, $\Lambda/v$ is in general not tenable. 

\section{The phase transition on the lattice}

In Table I we show all the points in the coupling constant
space for which Monte Carlo data was collected. In the simulations we
used two different programs checked against each other: one based on a
simple one-hit Metropolis algorithm and the other based on the hybrid
algorithm (with or without Fourier acceleration). The second algorithm
was always superior. Details concerning the codes and how they perform can
be found on the Appendix.
Following ref. \cite{YUE}, we have used as an order parameter the expectation
value of the $U(N)\times U(N)$ invariant operator
\be
O=\mbox{\rm Tr}\, {\bar{\phi}}^\dagger {\bar{\phi}}
\ee
which corresponds to the susceptibility, where
\be
\bar{\phi}_{ij}=\frac{1}{L^4}\sum_{x} \phi_{ij}(x).
\label{order}
\ee
The expectation value of the above operator is then proportional to
$v^2$ in the broken phase and zero in the unbroken phase, modulo finite
size corrections. We also measured the expectation value of
$O^\prime=\mbox{\rm Tr}(\phi^\dagger\phi)$ as an alternative order
parameter. 

We have used lattices of sizes ranging from $L=4$ to $L=14$. The
exact procedure we have used
depended somewhat on the region of ($\lambda_1, \lambda_2$) in which the
simulations were performed. In general, to obtain information
about the order of the transition at each given ($\lambda_1, \lambda_2$),
we have searched for hysteresis effects in the measurement of the order
parameter: we performed thermal cycles in the relevant parameter, $m^2$,
across the critical region where the field configuration from the last
run was used as input for the next run. Strong hysteresis loops are an
indication of a strong first order transition.
On smaller lattices, we have also looked at the histogram distribution
of  ${\rm Tr} \phi \phi^{\dagger}$, where a double-peak structure across
the critical region is an indication for two coexistent minima. This
procedure helps us identify the critical point of equivalent minima, as
well as the range of $m^2$ where metastability is observable. Tunneling,
of course becomes rare on larger lattices, which we have used to look
for coexistence. The measurements we provide in Table I for
the order parameter $v^2$ always refer to the largest lattice used.

\begin{figure}[htb]
  \epsfxsize 18cm \centerline{\epsffile{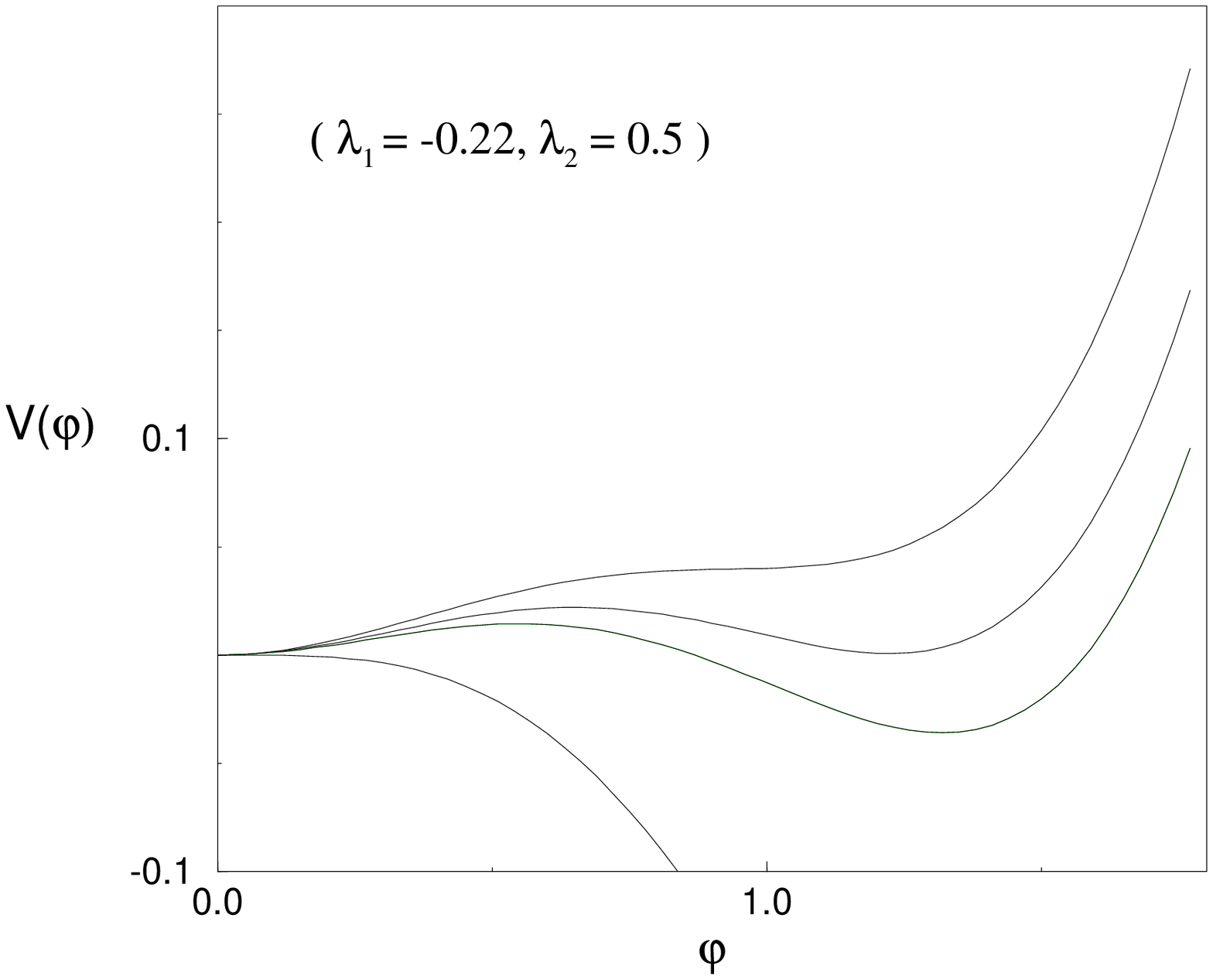}}
\caption{}
 \label{u_eff}
\end{figure}

Near a first order transition the effective potential develops two
minima. One of the minima, say $\langle \varphi\rangle=0$, is the lowest,
and as we increase the relevant parameter ($-m^ 2$) the nontrivial
minimum becomes dominant, and the system acquires a v.e.v.
$\langle\varphi\rangle=v\neq 0$, and eventually the minimum at the
origin disappears.  This does not imply that as soon as one of minima
becomes dominant the system jumps to it; near the transition there is a
potential barrier between them whose height determines the strength of
the transition and the tunneling rate.
The relation
\be
V''(v)\sim 1/\xi^2,
\label{rel}
\ee
tells us that the weaker the transition, the less steep the effective
potential will be, and the more difficult it will be to observe
metastable states.
Of course, if the correlation length near the transition is bigger than,
or of the order of, the lattice size itself, one should not expect to see
metastable states because the system is not able to see the
distinction between the two existing minima. In our simulations we have
used this property to get rough estimates of the correlation length.
Also, if the transition is weakly first order, it might happen
that one of the minima disappears very soon after the transition has
taken place, and the actual range of values of $m^2$
where metastable states  are detectable  is very narrow. A good deal of
fine tunning for $m^2$  is then called for.
All these features can be visualized by comparing figure \bref{u_eff},
corresponding to a relatively strong transition ($\lambda_1,\lambda_2)$
very close to the stability line, and figure \bref{u_eff2}, in which the
transition is weaker.

The order parameter of the transition, $v$, is the
quantity that, when expressed in physical units, gives, after gauging the
model, a mass to the $W^\pm$ and $Z$ bosons, according to the relation
\be
M_W={1\over 2} g v
\ee
However, for this relation to be true the residues of all
particles have to be properly normalized. This is not necessarily so
on the lattice and we are forced to distinguish between
$v_{phys}$, the physical value, and $v_{latt}$, the value we obtain
from our simulations. The relation between the two is
\be
v_{phys}=v_{latt}/\sqrt{Z_G}
\ee
where $Z_G$ is the Goldstone wave-function renormalization defined
by the unit residue condition.
The value of $Z_G$ has been estimated in \cite{KUTI} and
found to be always close to one. Although these results were derived at
$\lambda_2=0$, we expect that the wave function renormalization
$Z_G$ stays close to but smaller than one and thus taking it into account
can only increase the value of $v_{phys}$.

\section{Weak coupling}\label{WK}

\begin{figure}[htb]
  \epsfxsize 18cm \centerline{\epsffile{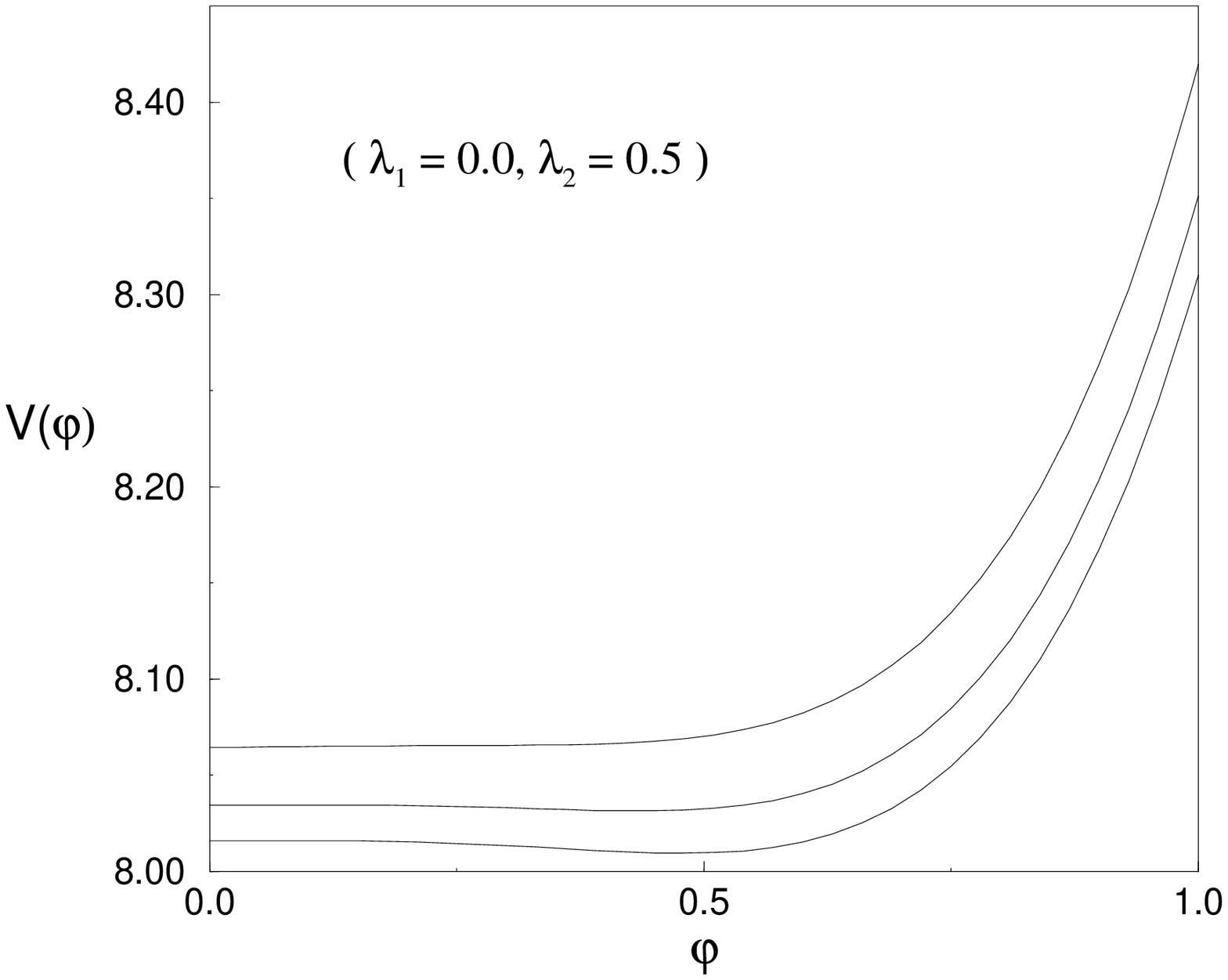}}
\caption{}
\label{u_eff2}
\end{figure}

In the weak coupling region ($\lambda_1,\lambda_2 < 1$),
lattice perturbation
theory does apply and can be used to compare with the numerical data.
At tree level (equivalent to assuming the mean field approximation) the
transition is always of second order. The symmetry breaking pattern
and the tree-level relations  have been described in section 2.
Radiative corrections change this behaviour.
The bare one-loop effective potential was computed in \cite{YUE}.  If
$\lambda_2>0$, the result is 
\bea
V(\vr)&=&\frac{1}{2}m^2 \vr^2 + (\lambda_1+\frac{\lambda_2}{2})\vr^4
+\frac{1}{2L^4}\sum_{p}\ln({\overline{p}}^2+\aam+12\vr^2(\lambda_1+
\frac{\lambda_2}{2}))
\nonumber
\\
&&
+\frac{3}{2L^4}\sum_{p}\ln({\overline{p}}^2+\aam+4\vr^2(\lambda_1+
\frac{3\lambda_2}{2}))
\nonumber
\\
&&
+\frac{4}{2L^4}\sum_{p}\ln({\overline{p}}^2+\aam+4\vr^2(\lambda_1+
\frac{\lambda_2}{2}))
\label{loop}	
\eea
where ${\overline{p}}^2=\sum_{\mu}2-2cos(p_{\mu})$ is the lattice
propagator. The quantity $\aam\equiv m^2-m^2_c$, where $m^2_c$ is
the value at which $V''$ vanishes at the origin, resums some two-loop 
corrections into the mass \cite{AMIT}.
On the other hand, if $\lambda_2<0$, one obtains
\bea
V(\vr)= &=&\frac{1}{4}m^2 \vr^2 + (\lambda_1+\lambda_2)\frac{\vr^4}{4}
+\frac{1}{2L^4}\sum_{p}\ln({\overline{p}}^2+\aam+6\vr^2(\lambda_1+
\lambda_2))
\nonumber
\\
&&
+\frac{2}{2L^4}\sum_{p}\ln({\overline{p}}^2+\aam+2\lambda_1\vr^2)
\nonumber
\\
&&
+\frac{5}{2L^4}\sum_{p}\ln({\overline{p}}^2+\aam+2\vr^2(\lambda_1+
\lambda_2))
\label{loopunstable}
\eea
The effective potential for the $(\lambda_1,\lambda_2)$ values
(-0.22,0.5) and (0,0.5), where perturbation theory should still be
valid, is shown in figs. \bref{u_eff} and \bref{u_eff2}.
As is manifest from these figures there are two coexisting minima,
hence the transition is of first order, albeit more weakly so as we
move away from the stability line, in accordance with the above
discussion of the Coleman-Weinberg phenomenon. Quantum corrections have
transformed the second order phase transition of mean field theory to
a first order one.

We now look at the numerical results in the weak-coupling regime and
compare them to  the one-loop potential results.
The predictions  from \bref{loop} hold, on average,  at the $10-30\%$
level for the values of  $(\lambda_1,\lambda_2)$ we have analyzed in the
weak coupling region, and should be more accurate away from the phase
transition region. It is indeed natural to expect that deviations are
indeed larger near the phase transition surface, at least when the
transition is weakly first order (as exemplified e.g. by
fig.~\bref{u_eff2}), since the precise location of the minimum of the
potential is in this case unstable against small corrections in its
shape originating from two-loop corrections and beyond.

\begin{figure}[htb]
 \centerline{\epsffile{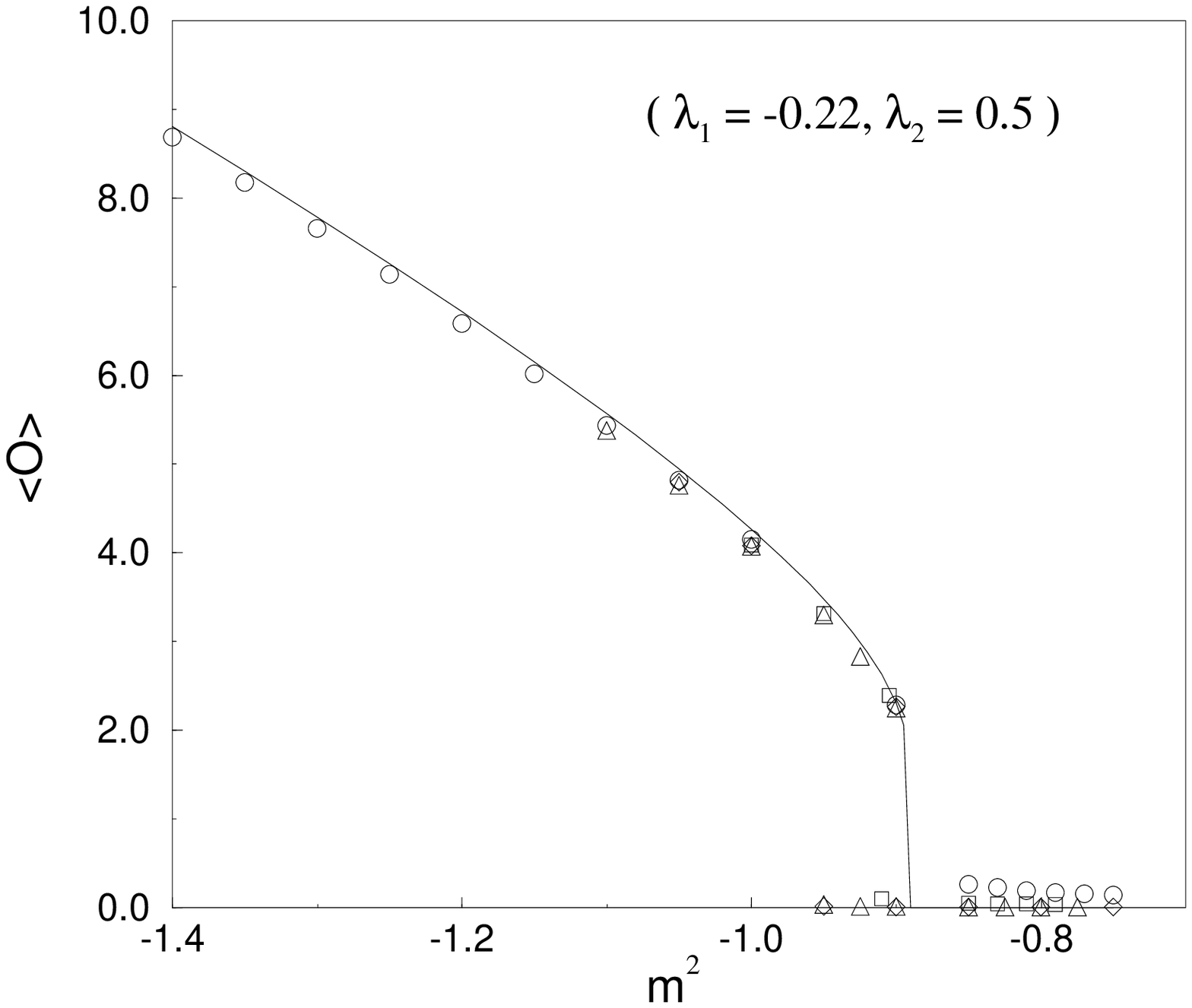}}
\caption{}
\label{ord225}
\end{figure}

\begin{figure}[htb]
  \epsfxsize 18cm  \centerline{\epsffile{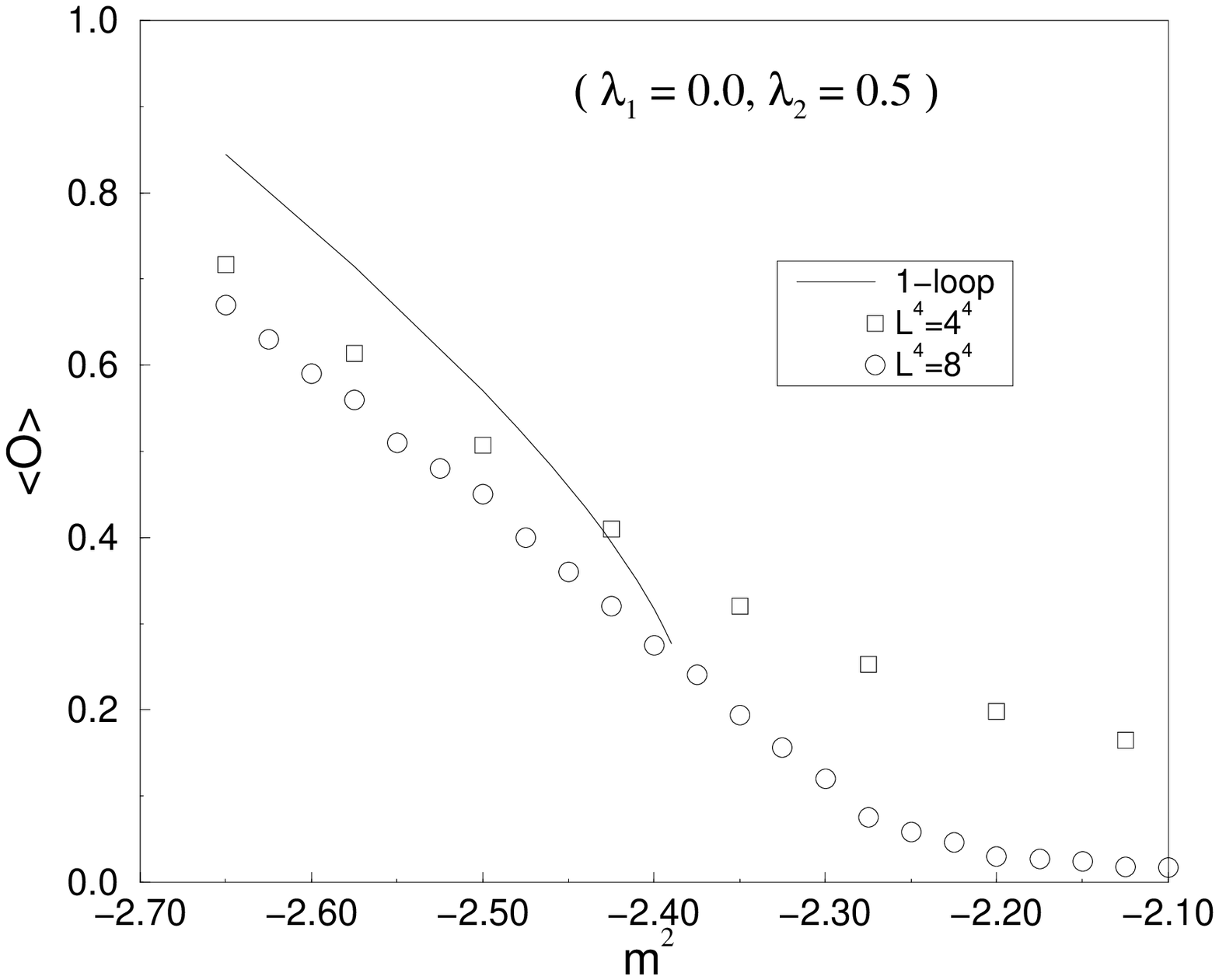}}
\caption{}
\label{ord005}
\end{figure}

In fig. \bref{ord225} we plot the results at $(\lambda_1=-0.22,
\lambda_2=0.5)$ where the correlation length was estimated to be $\xi
\sim 3$  from the effective potential. The smallest lattice size where
metastability was observed was for $L=6$. 
The transition  is  a relatively strong first order one.
Comparing the evolution of $v$ as a function of $m^2$ against
the effective potential prediction we see that the agreement is good.
Following our general argument we expect the transition at (0,0.5) to be
weaker since we are away from the stability line. The one-loop
effective potential calculation gives $\xi\sim 40$, so it is unlikely
that we can see metastability, even on our larger lattices. Also, we
expect the effective potential calculation to become less
reliable.  Fig. \bref{ord005} shows our data for the order parameter
compared to the one-loop effective potential on a lattice of the same
size. The agreement is certainly worse than before. The effective
potential still predicts a first order transition (at $m^2=-2.39$),
albeit a weak one. The jump in the order parameter $v$ is approximately
$0.71$. 

We have also analyzed the $(\lambda_1=0.5, \lambda_2=-0.45)$  point
where the symmetry breaking pattern is that of eq.~(\ref{SB2}). 
% Although this point is close to the stability line, we have found that
% the phase transition here is only weakly first order. 
The effective potential calculation suggests a first-order transition at
$m^2=-0.84$, where the correlation length is $\xi\sim 7$ and the jump the
v.e.v. is $v=0.91$. Our numerical
data agreed again within 30~$\%$ for the order parameter to these
predictions, although no hysteresis effects were observed even on the 
$14^4$ lattice.
% The perturbative renormalization group analysis suggests that $\Lambda/v
% \simeq 20, 36$ at one-loop and two-loop respectively, while 
% which is in agreement with our numerical results. 
Although points on the $\lambda_2<0$ region do not seem to correspond to
the phenomenological model of strong extended technicolor 
or top-condensation, based on the simple Nambu-Jona Lasinio model (in
the large $N_c$ color approximation), this needs not be the case in
general \cite{HASENFRATZ}. 

\section{Strong coupling}\label{SC}

The strong coupling region must be studied numerically.
The strategy we employed was the following.  We studied the smaller
lattices, $L=4, 6, 8$, using the hybrid algorithm usually, accummulating
about $10^5$ configurations. We searched for two minima in the
histograms corresponding to the expectation value of the operator ${\rm
Tr}(\phi  \phi^{\dagger})$.
We then moved to bigger lattices $L=12,14$ to look for coexistence.
Generally, coexistence was found on larger lattices for values of $m^2$
slightly more negative than on smaller lattices, due to finite size
corrections. 
Along the process of increasing the lattice  size we eventually begin to see
metastability at some size $L^*$. We estimate then the correlation length
to be $\xi \sim L^*/2$. Crude as this procedure may seem, it is physically
meaningful and it agrees, where comparison is possible, with the
effective potential.

\begin{figure}[htb]
 \epsfxsize 18cm \centerline{\epsffile{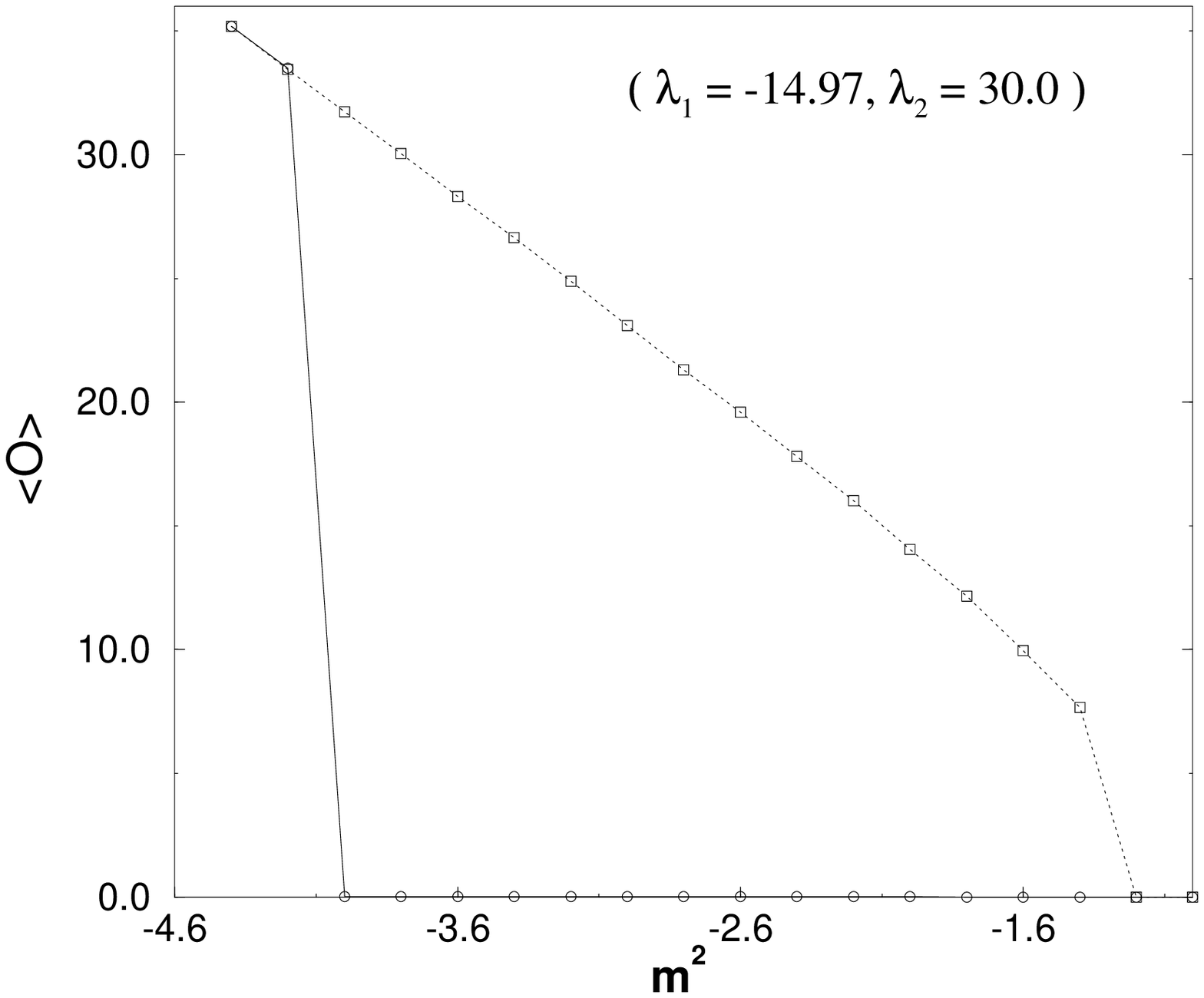}}
\caption{}
\label{hys3978}
\end{figure}

All points close to the stability line exhibited marked
hysteresis loops and hence show strong first order phase transitions.
As  an example we can take  the point (-3.97,8) where the corresponding
hysteresis loop is shown in fig. \bref{hys3978}. The transition becomes
stronger the upper we move along the stability line. Notice that since
$\xi\simeq 1$ the cut-off effects are big and the connection to 
continuum physics questionable. Similar conclusions apply to the point
($\lambda_1,\lambda_2$)= ($-14.97, 30$).

\begin{figure}[htb]
  \epsfxsize 18cm  \centerline{\epsffile{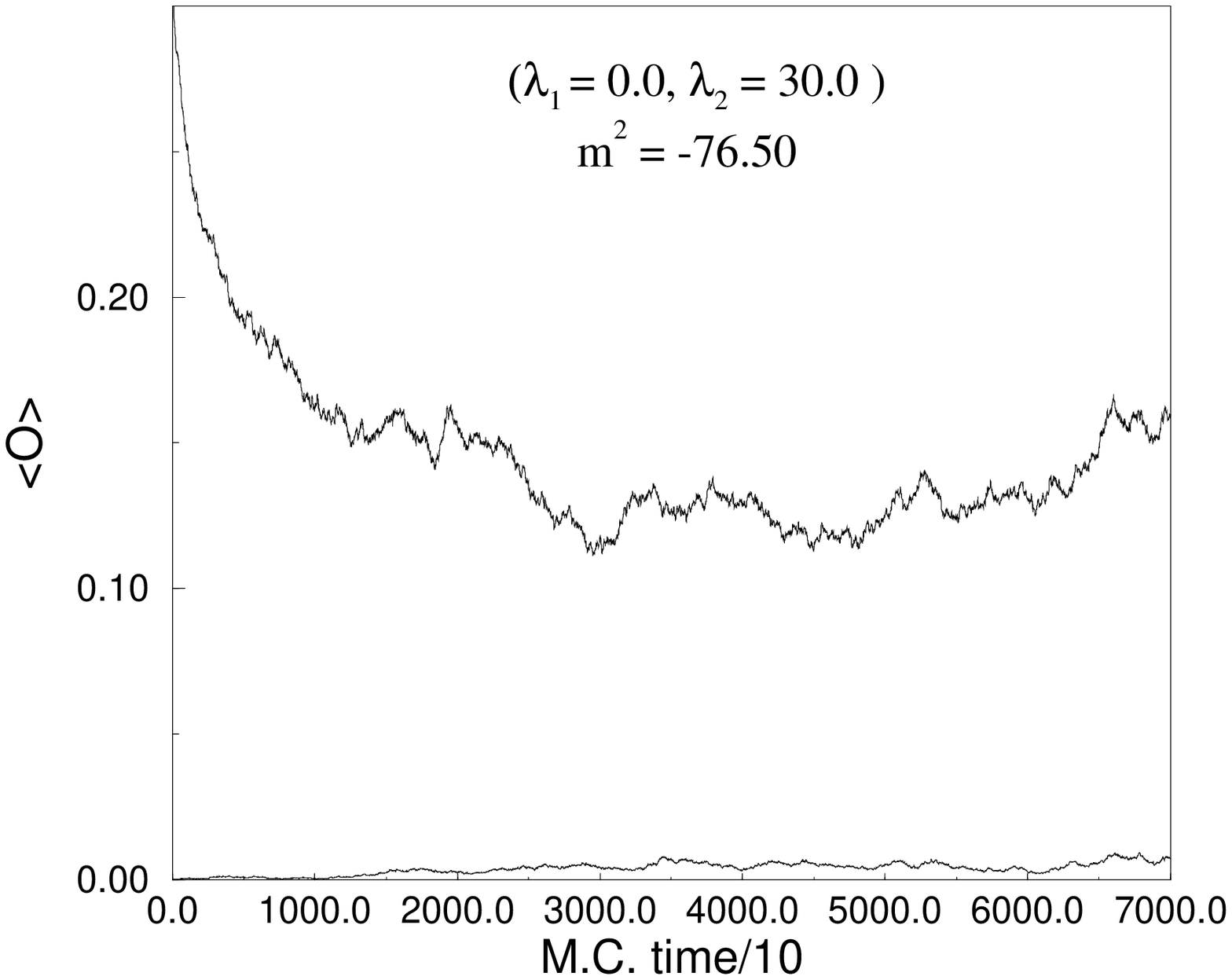}}
\caption{}
\label{cyc765}
\end{figure}

Points close to the $\lambda_1=0$ axis present always  weak
first order transitions. Typically runs on $L=4$ lattices do not
show  any hysteresis effects.  However we found
clear sign of the existence of two minima in  $L=12, 14$ lattices.  In
figure \bref{cyc2470} we display the  clear signal of two minima for the
point (0, 8),  and, similarly,  fig.  \bref{cyc765}  shows the two
minima signal for (0, 30). There is evidence that the transition is
stronger in the second case as the two signal minima can be seen in a
$L=10$ lattice. The transition gets indeed stronger with increasing
$\lambda_2$.

\begin{figure}[htb]
  \epsfxsize 18cm \centerline{\epsffile{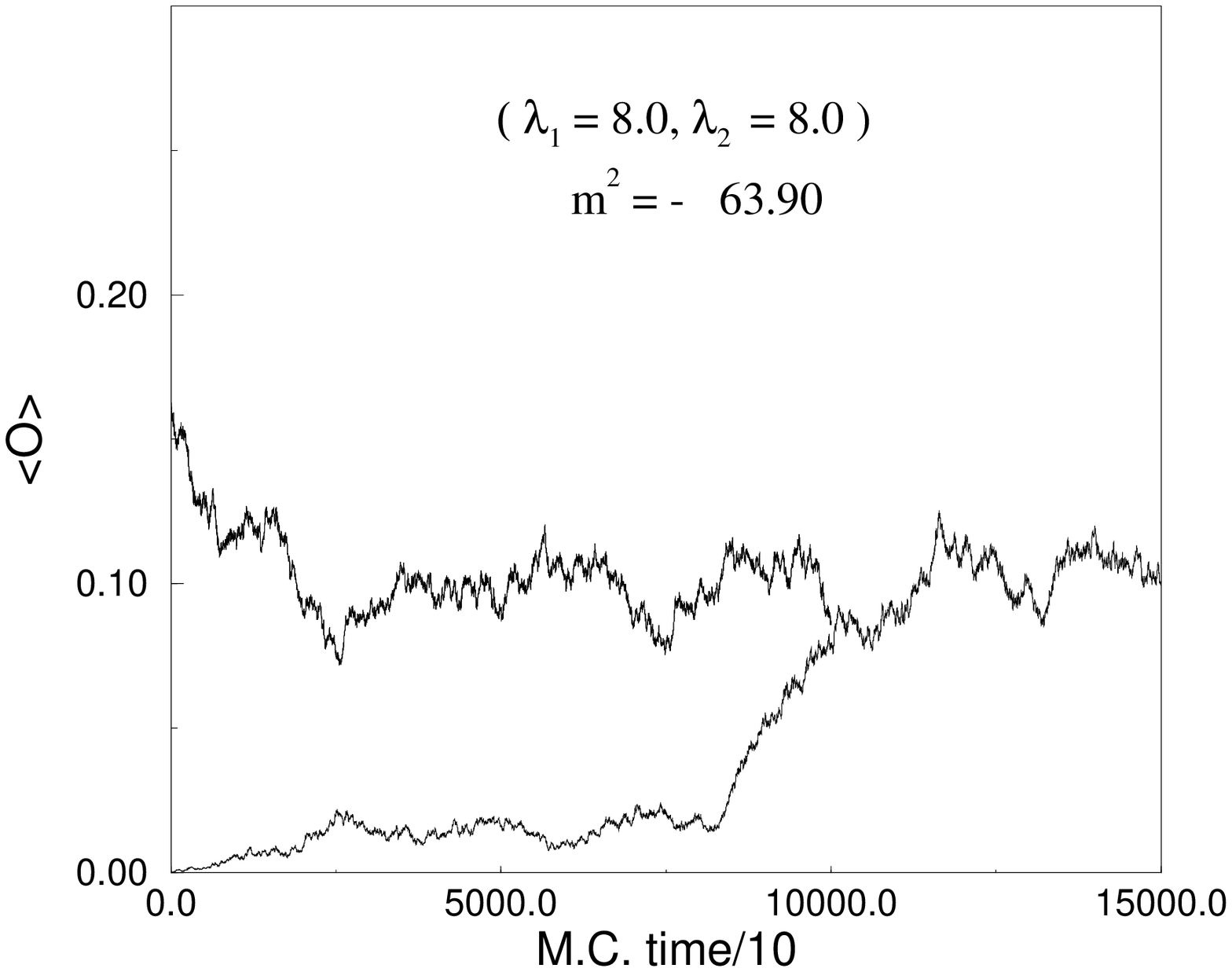}}
\caption{}
\label{cyc639}
\end{figure}

\begin{figure}[htb]
  \epsfxsize 18cm  \centerline{\epsffile{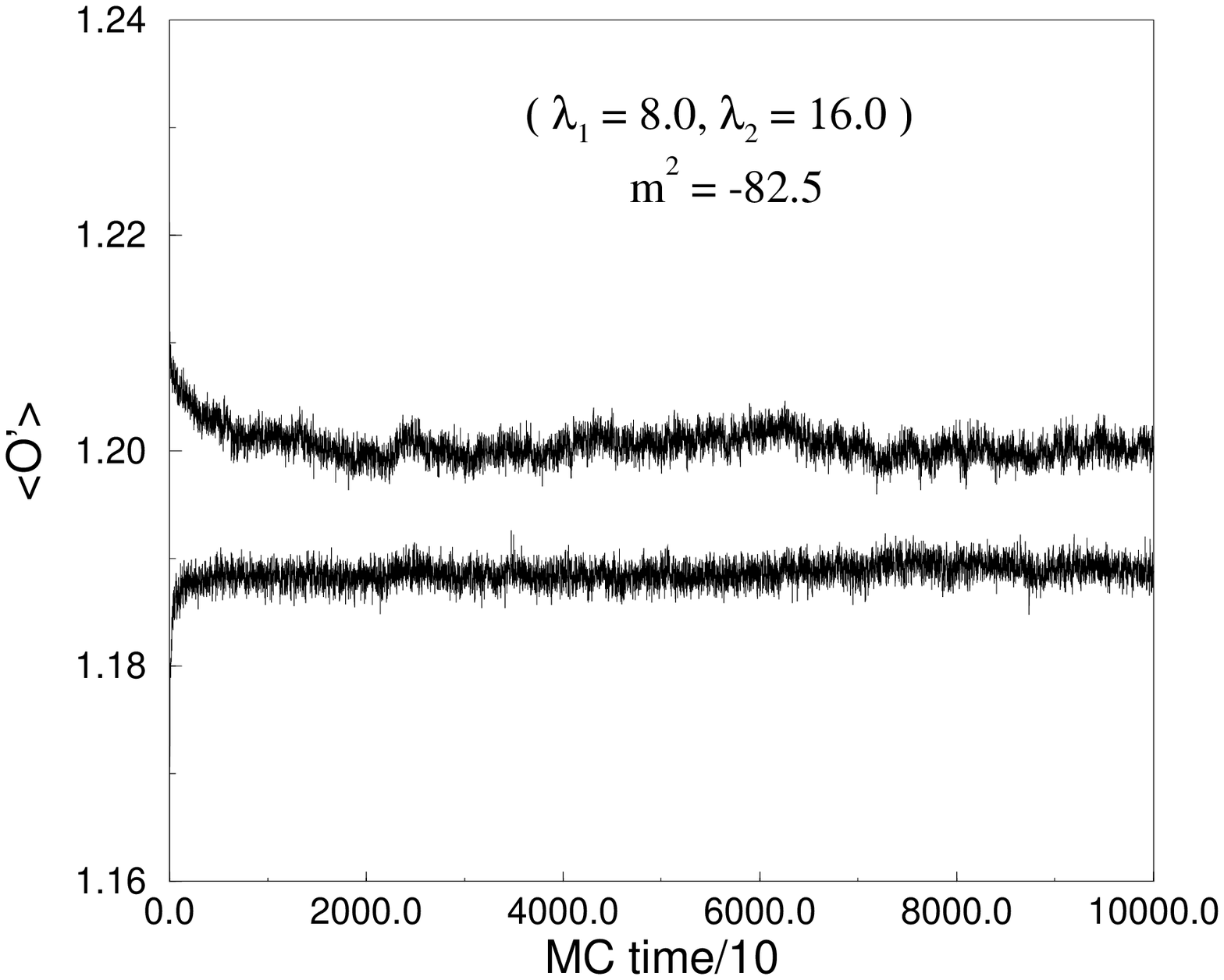}}
\caption{}
\label{cyc825phi2}
\end{figure}

For those points  deep in the $\lambda_1>0,\lambda_2>0$ region that we
analyzed, we were able to observe coexistence of phases, but only
in lattices of $L=14$. The transition is always clearly first order, but
characterized by correlation lengths much larger than those obtained close
to the stability line (this is of course as it should be, given the form
of the RG trajectories).
In figure \bref{cyc639} a plot is shown for  the point (8,8), where
the system eventually tunnels to the right minima. The symmetric phase
is in that case a relatively short-lived metastable state.
In fig.~\bref{cyc825phi2} we plot the Monte Carlo time evolution of the
operator ${\rm Tr} \phi \phi^{\dagger}$, starting with
ordered-disordered  initial conditions for the point (8,16).

For points close to the $\lambda_2=0$ axis, it is very difficult to
differentiate a weak first order from a second order transition.
More detailed methods with very high-statistics would be needed
\cite{ARN} complemented with finite-size scaling.

We have summarized the knowledge we have gained about the  value of the
order parameter at the transition and the corresponding correlation
length in Table~\bref{table}. From these results
we see that in most of parameter space (in the region where the
symmetry breaks the way we are interested in for phenomenological
reasons) the vacuum expectation value $v$ (at scale $\Lambda$, $v(\Lambda)$)
jumps to a value which is typically only one order of magnitude
smaller than the cut-off. The physically relevant v.e.v.  $v(v)$ (that,
after gauging, gives a mass to the $W^\pm$ and $Z$ bosons, and not
$v(\Lambda)$), according to the perturbative RG flows should be even
bigger.

\section{Conclusions}

In this paper we report an extensive Monte Carlo simulation
of the $U(2)\times U(2)$ model. We have found no evidence of the
existence of any fixed point other than the gaussian one at the
origin of the $(\lambda_1,\lambda_2)$ plane.

We have investigated many points in this plane using a variety of numerical
and analytical techniques.
We have been mostly interested in
getting a semi-quantitative picture of the symmetry breaking transition
over the different regions of the phase diagram in order to identify
regions of second or weakly first order transitions. There is no
evidence of any genuine second order transition, except if
$\lambda_2=0$.

For most of the $(\lambda_1,\lambda_2)$ values  in the
strongly coupled region, the  jump in the order parameter
parameter $v$ is approximately equal to $0.3-0.4$ in lattice
units. If we assume that the renormalization constant $Z_G$
is close to one, we can exclude the possibility
of a large hierarchy in that region.  We were aiming at values of
$v$ in the range $10^{-3}$, that is two orders of magnitude
smaller than the generic result.

We found just one region satisfying the requirement that the phase
transition is weak enough. This is $\lambda_2 \to 0$, the limit where
the model approaches the $O(N)$ linear sigma model. For small values of
$\lambda_1$, the tunning  in $\lambda_2$ must be extraordinarily
accurate, probably at the  $10^{-3}$ precision level or more. This is
evidenced by the effective potential calculations.
For larger values of $\lambda_1$ this is somewhat relaxed, as the
jump in the order parameter seems to increase
more slowly as we depart from the $\lambda_2=0$ line for a fixed
value of $\lambda_1$.
Phenomenologically viable models must then
lead to values  for the effective couplings which, at the
cut-off scale satisfy the above requirements.

All our data conforms perfectly with the standard picture
of first order phase transitions with runaway trajectories deduced from the
Coleman-Weinberg analysis. We have some evidence that the running is in some
cases very fast.

\bigskip

{\bf Acknowledgements}

We would like to thank Yue Shen for providing us some of his numerical
results, R.~Toral for multiple discussions and help concerning the use
of the hybrid algorithm, and J. Comellas for discussions. V.K. would
like to thank Sekhar Chivukula and Dimitris Kominis for many useful
discussions. A.T. acknowledges a grant from Generalitat de 
Catalunya. D.E. thanks the CERN TH Division for hospitality.
This work has been partially supported
by CICYT grant AEN950590-0695,
CIRIT contract GRQ93-1047 and by EU HCM network grant
ERB-CHRX-CT930343.

\bigskip

\newpage

\appendix

\section{Appendix: Algorithms}\label{app}
We have left  for this  appendix all the  technical details
of the numerical simulations. We have mostly employed the hybrid algorithm
since it
allows for a better control of the autocorrelation times and the rejection
percentage.

We consider the  generalized hamiltonian
\be
\bar{\ha{H}}(\phi,\pi)={\ha{H}}(\phi)+{\sum\pi^2\over 2},
\ee
where $\ha{H}$ is the hamiltonian of the physical problem at hand (in our
case $\ha{H}$ is just the euclidean action
of the $U(2) \times U(2)$ model), and $\pi$ some fake
 momenta conjugate to each variable $\phi$.
We especify some initial values for the momenta $\pi$ according to
a gaussian distribution, and then  numerically  integrate the Hamilton
equations
for the $(\phi,\pi)$ dynamical system.  Any algorithm can be used provided
that is
time-reversal and preserves the area of phase space\cite{HYB}. A convenient
way of  satisfying both requirements is to use the leap-frog algorithm
\bea
\phi(t+\delta t)&=&\phi(t)+\delta t  \  A\pi(t)
+\frac{(\delta t)^2}{2}\  A A^T F(t),
\nonumber
\\
\pi(t+\delta t)&=&\pi(t)+\frac{\delta t}{2}\  A^T
(F(t)+F(t+\delta t)),
\eea
where  $F=-\der{\ha{H}}{\phi}$, and $A$ is
some (arbitrary) $t$-independent matrix.
In the above expressions we use a vector notation for $\phi,\pi$ and $F$,
the vector index running over all lattices sites.
After a number of leap-frog steps, the resulting configuration
is subject to a standard Metropolis test. It can be either accepted
or rejected,  and in the latter case we start anew.
Using just one leap-frog step the hybrid algorithm would be strictly
equivalent to the Langevin one (the fake conjugate
momenta playing the role of the gaussian Langevin noise), except that
here we must pass the Metropolis test, which makes the algorithm
exact. In general it will be convenient to use several
leap-frog steps before attempting the Metropolis test.

We have tried two different choices for the matrix
$A$:  the identity, $A=I$ (standard hybrid algorithm (SH)), and
\be
A_{n,m}=\frac{1}{L^d}\sum_{p} \exp -ip(n-m)\varepsilon(p)
\ee
where $\varepsilon(p)= (\overline{p}^2+m^2)^{-1}$ is the free
lattice propagator. The latter correspond to the Fourier accelerated
hybrid algorithm (FA)\cite{HYB}, and it allows for
an update of all modes with similar efficiency. This, combined with the
decorrelation induced by the numerical integration, makes for a very
robust algorithm as far as beating critical slowing down goes. However,
due to the need of performing  fast Fourier transforms in four
dimensions, FA  is intrinsically much slower than  SH. The gains
in beating critical slowing down are only apparent  for large
correlation length.
However, at ($\lambda_1, \lambda_2) = (-0.22,0,5)$
in a $8^4$ lattice, the autocorrelation time at $m^2=-0.895$ (transition
point) is about four times bigger for the SH than for FA, making FA
useful but not really necessary. This is perhaps not too surprising
since the correlation length is in much of the parameter space
relatively small, even close to the transition. 

Two parameters have to be adjusted in the hybrid algorithm, namely the
number of leap-frog steps and the step size $\delta t$. They are the
equivalent to the fudge parameter one uses in a standard Metropolis to
adjust the acceptance rate.
If we use a relatively large step size $\delta t$,  successive
configurations soon become more uncorrelated.  However a large step will
decrease the acceptance rate, so a compromise must be reached. We have
done extensive tests in the simple case $\lambda_1=0, \lambda_2=0$,
which can of course be solved analytically. The best situation
seems to be to take $\delta t$ such that the acceptance rate is about
$90\%$. On lattices ranging from $8^4$ to $12^4$ this
corresponds to taking $\delta t\sim 0.2$ (depending somehow on the
values of ($\lambda_1, \lambda_2$).
The other freedom concerns the number
of leap-frogs before the rejection Monte Carlo is performed. The larger
the number of leap-frogs, the smaller the autocorrelation time,
but the required computer time  increases too and, at some
point, nothing is gained by decorrelating even less our observables
(providing the rejection rate remains low).
In our case, the optimal choice for  the $8^4,12^4$
lattices were  between 5 and 7 leap-frog steps.

After taking all these precautions, the hybrid algorithm
works remarkably well. As a check we have verified that
we are able to reproduce the results for the free theory with
very high accuracy. On a $L=8$ system, with $m^2=1$, it is not
difficult to get after ${\cal O}(10^6)$ Monte Carlo steps
four or five significant figures.

The algorithm seems to work efficiently for all the
$(\lambda_1,\lambda_2)$ values we have tested.  For comparison
we have written a conventional Metropolis Monte Carlo code.
Not surprisingly, the improvement brought about by the
hybrid algorithm depends substantially on the correlation length.
When the phase transition is clearly first order, Metropolis and
hybrid fare similarly (the latter being about twice as fast). Hybrid
gets better when the correlation length grows.

\vfill
\eject

\newpage
\section{Table}

%  Table
%

\begin{table}
\begin{center}
\begin{tabular}{|c|c|c|}     \hline \hline
$(\lambda_1,\lambda_2)$ &  $v^2$ & $\xi$ \\ \hline
(0.5, -0.45) & 0.83  &  7 \\ \hline

(-0.22, 0.5) & 2.10  &  3  \\ \hline

(-3.97, 8)   & 10-20      & $<$ 2   \\ \hline

(-14.97, 30) &  20-40     & $<$ 2   \\ \hline

(0, 0.5)     &  0.5  & 40   \\ \hline

(0, 8)       &  0.11     & 6   \\ \hline

(0, 16)      &  0.09  & 6 \\ \hline

(0, 30)      &  0.15  & 6    \\ \hline

(8, 8)       &  0.11   & 6    \\ \hline

(8, 16)      &  0.16   & 6   \\ \hline

(8, 30)      & 0.16   & 6  \\ \hline\hline
\end{tabular}
\caption{Estimates of the jump in the order parameter $v^2$ and
 correlation length estimated by $\xi\simeq L^*/2$, where $L^*$ is the
 smallest lattice where coexistence was found, or from the effective
 potential.} 
\label{table}
\end{center}
\end{table}

\newpage
\section{Figure Captions}

\begin{figure}[htb]
  \epsfxsize 18cm  \centerline{\epsffile{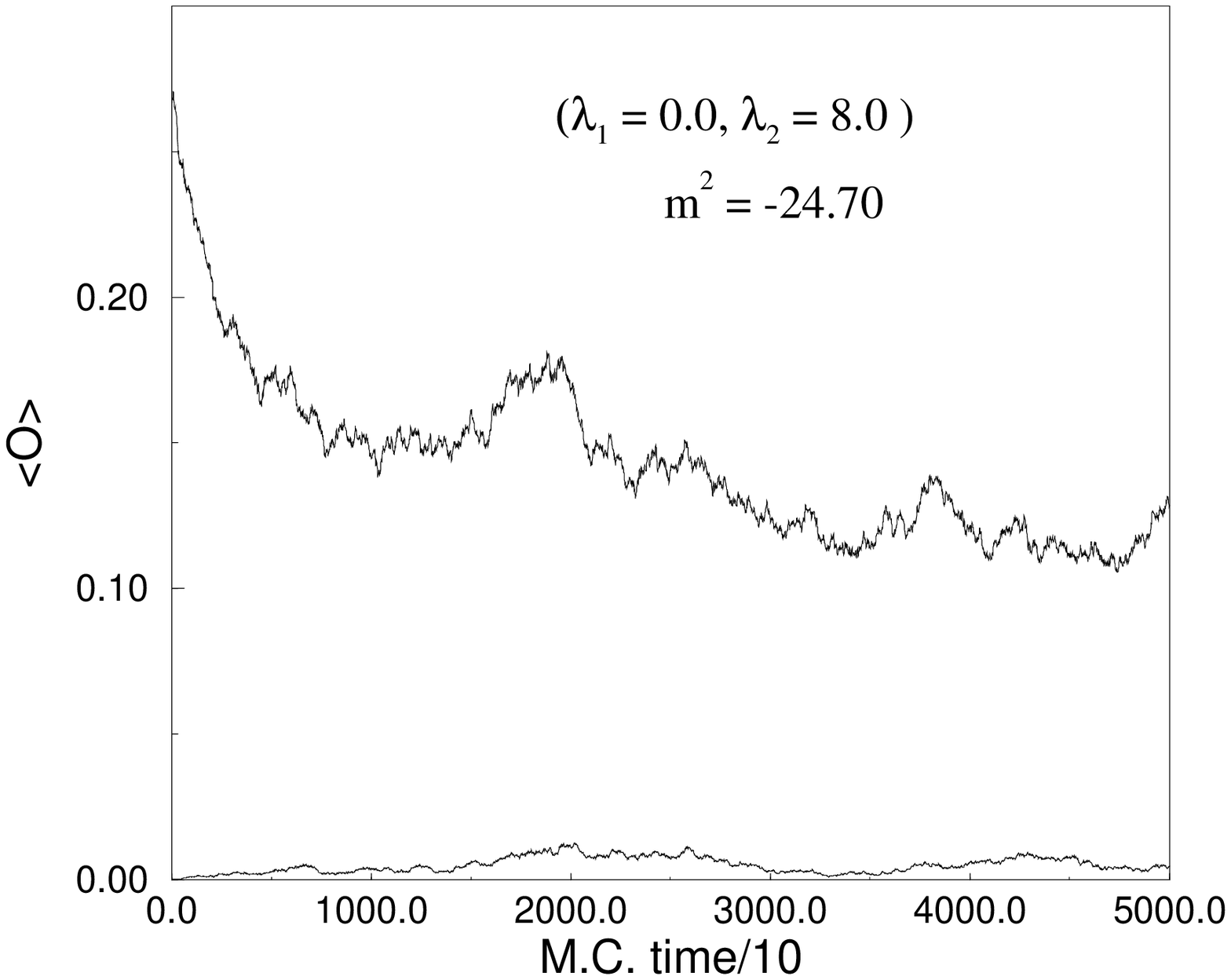}}
\caption{}
\label{cyc2470}
\end{figure}

\begin{itemize}
\item
Figure \bref{traj}.-
Perturbative RG trajectories for the $U(2)\times U(2)$ model starting
from bare couplings ($\lambda_1(\Lambda), \lambda_2(\Lambda)$) along the
line $\lambda_1=2$ or $\lambda_2=2$. The solid lines (dash lines)
correspond to one-loop (two-loop) trajectories while the stability line
is indicated as a dotted line. 
Indicatively, the dots along a trajectory represent the evolution of the
couplings after running by a factor of $e$ down to the infrared. 
Notice that while  for bare couplings with $\lambda_2<0$ or close to the
stability line the flow is to the left towards the stability line,
there is a region with $\lambda_1,\lambda_2>0$ where the flow
is reversed and it appears that it never crosses the stability line.

\item
Figure \bref{u_eff}.-
The effective potential for different values of $m^2$
at $(-0.22,0.5)$. The effective potential is arbitrarily normalized
so that $V(0)=0$. The four lines correspond  to
$m^2=-0.88$ (ordered phase), $m^2=-0.90$ (broken phase is metastable),
$m^2=-0.94$ (symmetric phase is metastable) and
$m^2=-1.12$ (disordered phase).

\item
Figure \bref{u_eff2}.-
The effective potential for different values of $m^2$ at $(0,0.5)$.  We
have shifted the origin for the different $m^2$ values in order to be
able to visualize it. 
The values of the mass are $m^2=-2.37$ (ordered phase),
$m^2=-2.42$, and $m^2=-2.45$. The important point to note is how weak
the barrier that separates two minima is. 

\item
Figure \bref{ord225}.-
The expectation value of order parameter $<O>$ as a function of $m^2$ at
$(\lambda_1,\lambda_2)=(-0.22,0.5)$. 
The solid line is the one-loop prediction; circles correspond to a $L=4$
lattice, squares to  $L=6$ lattice, triangles to   $L=10$,  and diamonds
to $L=12$.

\item
Figure \bref{ord005}.-
The expectation value of order parameter $<O>$ as a function of $m^2$ at
$(\lambda_1,\lambda_2)=(0,0.5)$. The solid line is the one-loop
prediction, squares correspond to $L=4$, and circles to $L=8$.

\item
Figure \bref{hys3978}.-
Plot of the hysteresis loop near the classical stability line at
$(-3.97,8.0)$. The results correspond to a $L=4$ lattice with
$10^5$ configurations after thermalization.

\item
Figure \bref{cyc2470}.-
Time history of two runs starting with ordered and disordered
conditions, respectively,  at $m^2=-24.7$
in a $L=12$ lattice, clearly displaying the two minima signal.
The point in parameter space is (0,8).

\item
Figure \bref{cyc765}.-
Same as in fig. \bref{cyc2470} for $m^2=-76.5$ at point
$(0,30)$. The lattice size is $L=12$.

\item
Figure \bref{cyc639}.-
Same as in fig. \bref{cyc2470} for  $m^2=-63.9$ at point (8,8). Size is
$L=12$. At some point the system jumps to the broken phase.

\item
Figure \bref{cyc825phi2}.-
Evidence for coexistence at the point (8,16)
and $m^2=-82.5$ in a $L=12$ lattice. The operator we plot here
is not the square of the order parameter as previously, but rather
$O^\prime={\rm Tr} (\phi \phi^{\dagger})$.

\end{itemize}

\newpage

\section{figures}

% \begin{figure}
% \centerline{\psfig{figure=ord225.eps}}
% \caption{}
% \label{ord225}
% \end{figure}

% \begin{figure}
% \centerline{\psfig{figure=fig5.eps}}
% \caption{}
% \label{ord005}\end{figure}

% \begin{figure}
% \centerline{\psfig{figure=hys3978.eps}}
% \caption{}
% \label{hys3978}
% \end{figure}

% \begin{figure}
% \centerline{\psfig{figure=cyc247.eps}}
% \caption{}
% \label{cyc2470}
% \end{figure}

% \begin{figure}
% \centerline{\psfig{figure=cyc765.eps}}
% \caption{}
% \label{cyc765}
% \end{figure}

% \begin{figure}
% \centerline{\psfig{figure=cyc639.eps}}
% \caption{}
% \label{cyc639}
% \end{figure}

% \begin{figure}
% \centerline{\psfig{figure=cyc825phi2.eps}}
% \caption{}
% \label{cyc825phi2}
% \end{figure}

\end{document}